\documentclass[]{article}

\usepackage{hyperref}
\usepackage{amsfonts}
\usepackage{amssymb}
\usepackage{amsmath}
\usepackage{color}
\usepackage{natbib}
\usepackage{setspace}

\begin{document}

\begin{center}
	{\centering{\huge\bf\it Data learning from big data}}\begin{center}
		
	\end{center}
\end{center}


\begin{center}
{\Large Jos\'e L. Torrecilla} \\
Institute UC3M-BS of Financial Big Data. Universidad Carlos III de Madrid, Spain\\joseluis.torrecilla@uam.es\\ \vspace{20pt}
{\Large Juan Romo}\\
Department of Statistics. Universidad Carlos III de Madrid, Spain
\end{center}



\begin{abstract}
Technology is generating a huge and growing availability of observations of diverse nature. This big data is placing {\it data learning} as a central scientific discipline. It includes collection, storage, preprocessing, visualization and, essentially, statistical analysis of enormous batches of data. In this paper, we discuss the role of statistics regarding some of the issues raised by big data in this new paradigm and also propose the name of {\it data learning} to describe all the activities that allow to obtain relevant knowledge from this new source of information.
\end{abstract}


\section{Introduction}
Big data is one of the most fashionable concepts nowadays: everybody talks about it, is permanently in the media, and companies and governments try to exploit the new amount of available information \citep{lohr2012,john2014,james}. The ideas behind this interest are mainly two. First, the fact that at present, most activities generate data (with very low cost) that contains (potentially valuable) information. The second one is well summarized in \cite{john2014}: ``Data-driven decisions are better decisions - it is as simple as that. Using big data enables  managers to decide on the basis of evidence rather than intuition.'' The opportunities offered by big data are undeniable, but there is still a debate about the scope and usefulness of this \cite{secchi,bul}. 		The opinions of the most fervent followers speak of the end of the theory and the models and, in articles like the controversial ``The end of theory'' \cite{end} they argue that ``with enough data, the numbers speak for themselves''. On the other hand, there have been more critical voices that question whether the optimism and the faith that is being put into the big data are really justified. In this line, Tim Harford wonders if ``we are making a mistake'' in another provocative article \cite{mistake}. In this paper we will review some of the big data aspects that can generate doubts from the point of view of a statistician trying to scrutinize if the data are sufficient by themselves or it is necessary to give them a sense.

First, it is convenient to be more specific. Although there is no single definition, there seems to be a certain consensus that big data encompasses the study of problems so ``Big'' that conventional tools and models can not handle them, either because they are not adequate or because they require too much time. In any case, whatever the definition we choose or where we put the emphasis, what is clear is that current technology generates huge amounts of data, so we have to be able to extract the best information from them and use it to make the best decisions. How to get it and the challenges associated to this new framework have become common discussion topic in the last years \citep{lynch2008,fan2014,gandomi2015} and the best way to tackle the problem has also been subject of debate. As an example, we can cite the former paper by Breiman \cite{breiman2001} about the two cultures of statistical modeling: stochastic models and algorithms (see \cite{dunson} for recent discussion in the context of big data). In what follows, we discuss the role of statistics regarding some of the issues raised by big data in this new paradigm and also propose the name of {\it data learning} to describe all the activities that allow us to obtain relevant knowledge from this new source of information.

From the classical statistics point of view this massive amount of data could be a blessing as we should be nearer to the real populations and the asymptotic convergence of the models. However, in practice, it looks more like a curse. Big sample sizes entail storage and processing problems and a huge effort must be done in terms of improving computational performance and developing new tools (both hardware and software) in order to handle such volume of data. But the real challenge for statisticians is to deal with the heterogeneity inherent to big data which appears in different forms. Populations are no longer homogenous or normal, consisting of different groups and communities, what makes invalid (at least partially) most classical statistical approaches based on convergence and central limit theorems. But heterogeneity also appears in the variety and complexity of the data managed in the areas where this kind of problems appears, such us medicine \citep{bio,health} or business \citep{management}. Therefore, the data at hand are far from standard random vectors in $\mathbb{R}^n$ stored in simple matrices. On the contrary, data can be completely unstructured coming from surveys, call records, web activity, social networks, an so on. All of this opens the door to a wide variety of problems including a proper information encoding and the combination of different types of data structures (categorical and continuous variables, functional data, time series, images, trees, text, networks, video...), some of them so recent and complex that they are a new field of study in statistics by themselves \cite{marron2014}. Moreover, big data tends to include (and often increase) the usual problems of high dimensionality. So we are in front of a double curse in both the number of features and observations.

Hence, the ``Big'' in big data refers, at least, to three different directions: velocity, complexity (number of variables and variety of data structures) and sample size. Usually these three problems appear together, but they entail different issues and the role of statisticians is quite different at each of them.


\section{Velocity}

Nowadays, we are able to generate such huge amounts of data (structured, unstructured and semi-structured) that would be very costly and take too much time to analyze them with conventional methodologies. Therefore, it is necessary to develop new tools and architectures in order to process the increasing and unstructured volume of data in a proper way. This represents a paradigm shift that affects all aspects of data processing that must be reconsidered \cite{tecnico}. 

Maybe, the best known indicators of these changes are the transition from sequential to parallel/cloud computing and the emergence of non-relational data bases. Substantial efforts have already been dedicated to parallelization at all levels. Platforms like Amazon Web Services, Azure or Google offer big data services and provide the infrastructure ``in the cloud'' on-demand and with elastic load balance, what solves many of the problems related to scalability and the maintenance of a physical cluster \cite{cloud}. Beyond the cloud computing, softwares like Cloudera or Hortonworks provide a wide catalog of parallelized services (storage, pipelines, data bases, data processing) and control the communication among them. Most of this software is based on Apache Hadoop (2008), a software framework that allows distributed storage and data processing using MapReduce. Finally, the cluster-computing framework Apache Spark (2014) is gaining popularity and substituting Hadoop for certain tasks. One of the reasons in favor of Spark are the programming interfaces for languages such as Python or R, largely used in machine learning and statistics \cite{spark}. 

Furthermore, non-relational or NoSQL databases arise from the need to both treat and search the data quickly and organizing the unstructured information. These models go a step further the parallelization of conventional relational databases (which is possible using Hive or HBase) by proposing new approaches that allow for greater dynamism and facilitate maintenance and scalability.

We could go deeper into these points, which are of great interest, and talk about other technical difficulties related to velocity (for example, in visualization or with the consistency of data), but the central task of statisticians here is, firstly, the development of computationally efficient algorithms. This aspect, which is usually forgotten or overlooked in the statistical literature, is critical in this context because even an optimal algorithm is completely useless if it cannot be applied.

\section{Complexity}

\pagebreak
Mostly, the complexity inherent to big data comes from the high dimensionality of the observations and the unstructured nature of the data we generate with smart phones, sensors, social networks, internet searches, GPS devices, emails, and so on. 

Problems associated with the dimensionality are well known by statisticians and other researchers. They have been extensively studied and many proposals have been done.  Beyond the particular properties of each technique, the dimension reduction methods can be grouped in two big families that we could call projection data and variable selection. 
Both approaches (variable selection and projection methods) have been extensively studied and compared at different contexts. The difference between these two approaches is the role of the original variables in the reduced space. While variable selection is restricted to the original variables, projection methods allow certain combinations of them in order to obtain the new components. This makes variable selection techniques to provide meaningful reductions (highly appreciated in areas such as biology or medicine).

Both methodologies have proven to be effective in a wide range of problems and are already being applied to big data questions where the issues typically associated with the high dimension take on special relevance. For example, \cite{singh2014} uses random forest in a distributed framework for variable selection, an evolutionary algorithm based on MapReduce is proposed in \cite{peralta2015evolutionary} for big data classification and \cite{bolon2015} provides a survey of some feature selection algorithms for big data problems ranging from DNA microarray analysis to face recognition. Furthermore, parallel penalized coordinates descent methods as those used for lasso optimization and others are studied in \cite{richtarik2016parallel}, and fast versions of other traditional algorithms have been proposed, as this version of PCA based on randomization \cite{abraham2014fast}. Finally, reduction methods applied to more complex structures can also be found, for example subgraphs \cite{graphs}.

An important example of big data with high dimensional observations are functional data. Functional Data Analysis (FDA) deals with objects of infinite dimension and problem related with this functional nature are sometimes similar those associated with big data (see, e.g., \cite{ahmed2017,goia}). In particular, a very relevant question in functional data analysis is dimension reduction (see \cite{vieu}).

Despite the number of works about these topics, there is still room for research. Most of these proposals just apply parallelization tools (as those commented before), stochastic search or on-line learning techniques aiming at accelerating existent algorithms. This computational efficiency is necessary, and it is worth to explore new searching strategies, matrix processing techniques, and so on. In fact, the mere parallelization of existing methods is not trivial since some very basic functionalities are not easily parallelizable (see, e.g., the median, which is not the median of the medians). Otherwise, other aspects can be also considered in order to answer natural questions like: What are the relevant variables in this heterogeneous context? How can we find and compare them? How to deal in a non trivial way with unstructured information? 

These and other problems are largely motivated by the variability and characteristics of the observations in big data problems (not just the dimension). Here, there is no longer a set of homogeneous measures, or even numeric items. The new variables can come from very different contexts and we do not even know how to code them adequately to extract the useful information. We must explore new ways to develop effective indicators, starting, for example, from opinions in social networks or from telephone calls to understand problems such as the level of satisfaction or recommendation systems. The complexity of some structures means that in some cases we lack even basic descriptive statistics and so it is necessary to propose new metrics or ways to establish relationships in order to be able to measure and compare \cite{marron2014}. Finally, it will also be important to study the properties of new methods and measures and to develop new visualization tools. \cite{visualization}.

\section{Big samples}

We have been dealing with the dimension problem, even with certain levels of heterogeneity, for many years and we have some experience to face them. Nevertheless, big data has brought a series of new problems related to the very high number (and often, low quality) of observations that were not expected a few years ago. On the contrary, from a classical point of view, having a lot of data was always considered as something positive that would make the models converge and practically allow us to achieve population results. We have often read and said phrases like ``the problem is that we do not have enough data'' and explain to our students how having enough data makes everything work. Well, now we have a lot of data and not everything works; in fact, what was promised as a blessing is rather a kind of curse with unexpected consequences that open new lines for research. Below we will briefly discuss some of the problems that have arisen due to the huge size, and sometimes low quality, of the samples.

\paragraph{Heterogeneous sources}
In big data applications, it is no longer common for observations to come from a single source with a single manageable coding. Quite the contrary, what we are usually interested in is capturing data from different sources  (with different levels of information, preferably complementary) which, unfortunately, provides samples collected in very different ways \cite{gandomi2015}. Let's think, for example, in a big company. It will probably be interested in its expenses and revenues, costs and benefits, but also in the performance of the last advertising campaign, in the relationships between its clients, in its public image or in levels of satisfaction among its clients. Data related to some of these topics will be obtained in a traditional and structured way (tabulated) but it will be also necessary to analyze comments on social networks, customer calls, survey results, etc., and include, as far as possible, exogenous variables such as demographic and economic indicators. Hence, this whole process combines structured, non-structured and semi-structured information from which one have to create comparable variables that capture the really useful information. Therefore, it is necessary to combine sources efficiently and 'consistently'. Currently, companies are already doing it, with more or less success, at purely heuristic levels but we can look for a sense in these combinations.

\paragraph{Subpopulations-clusters}
Basic hypotheses when working with statistical inference models are independence and identical distribution of the data. Moreover, there exist models for dependent data and ways to detect subsamples with different distributions. However, what in conventional statistics are particular cases, in big data problems becomes the general rule \cite{fan2014}. Since reality is complex and big data capture big pieces of this reality, it seems natural to have complex distributions including mixtures of different populations. This leads to new settings, starting with the failure of traditional models. Having several subgroups introduce difficulties in the study and analysis of the data, such as the Yule-Simpson effect. It seems interesting to find the different groups that exist in the sample and characterize them for their study, perhaps applying different models to different groups. Then it is relevant to decide what to do with the minority groups that are no longer so, are they outliers? Should we ignore them? Correct them? Study them separately (in some problems can be groups of special interest)? Otherwise, since we have mixtures, we have no longer normality with all its consequences.

\paragraph{Representation}
In spite of the volume of data from different subgroups, we found several problems with the representation of the sample. This is not due to the number of data but to the quality of the source which can be not representative of the population. Let's think about a customer service, probably the most likely customers to fill out a survey or answer some questions about the service will be the most satisfied and, above all, the most disappointed. This type of behavior is clearly reflected on the Internet and social networks, resulting in extremely skewed and noisy samples (missing values, lies, errors,…) \cite{elliott2017}. At this stage, it seems appropriate to wonder how to evaluate the representativeness of the sample and remove the non-probabilistic sampling bias, how to estimate the non recorded individuals or how to clean missing values and different types of wrong data.

\paragraph{Significance}
Paradoxically, in big data everything is significative. If the number of observations is big enough, everything is significant in practice even though it is not really the case.  This is happening, for instance, in the appearance of a multitude of spurious correlations or  the regression context where many (even all) variables are marked as significant by the standard criteria, but they are completely irrelevant for the model \cite{fan2014,gandomi2015}.

\paragraph{Nonparametrics}
The data speaks to us, but it is not always easy to understand in what language they are doing it. Making assumptions can greatly simplify this problem, but in a context in which it is difficult to hypothesize about the distributions of the data, it could be too limiting. On the other hand, nonparametric methods are free to learn from the sample without strong assumptions and therefore, limitations (at least theoretically). However, some care needs to be taken when using them. Nonparametric methods tend to be slow and having a lot of data, the time cost can be excessive. They are also prone to overfitting. Another problem that especially affects these techniques, and that is emphasized in big data settings, are unbalanced samples. In many problems of interest, we are concerned about finding/modeling a very small subset of the population (people with a certain disease, potential delinquents, rare but expensive events,…). The effect of this type of samples is sometimes dramatic and measures must be adopted to minimize it as much as possible, by generating balanced samples, cost functions or weights.

\paragraph{Static versus dynamic}
Statistics tend to be static. We have a sample, we use this sample to estimate our model which we apply to data that we assume to be similar to those we have used in the estimation. However, we have already seen that homogeneity is conspicuous by its absence. We have also seen that another feature of big data is the velocity of acquisition of new data.This large volume of incoming data can modify the sample so that the models we use are no longer valid (something already common in areas such as finance). These eventualities should be considered by exploring techniques to keep the models up-to-date, such as continuous (on-line) learning or models with the capacity to evolve according to the new incomes.

\paragraph{Semi-supervised learning}
Until recently, when we spoke of statistical (or automatic) learning, we used to consider two main branches: supervised and unsupervised. That is, labelled and unlabelled data respectively. Now, a third way is gaining importance, the semi-supervised learning. It is based on the impossibility (so expensive) of labelling all the data but at the same time being able to enjoy the advantages that brings, sometimes, have labels. A clear example is the classification of images, perhaps the most popular of the semi-supervised problems nowadays. 

\paragraph{The good versus the optimal}
One of the aspirations of a statistician is to find the optimal method to solve a problem. However, given the characteristics of this new paradigm, it seems difficult (at least at the present time) to find viable optimal procedures (that adjust to the real data and executable in reasonable time). Perhaps, it is time to partially sacrifice this idea and search for optimal approaches given certain circumstances. In addition, statisticians can help to control that the tools that are being developed (usually heuristic) make sense.

\paragraph{Trade-offs}
This is a world of equilibria. We have considered the balance between the good and the optimum and we mentioned the debate on data science versus modelling (which has been brilliantly treated by leading researchers like Breiman \cite{breiman2001} and, more recently, Donoho \cite{donoho2015}), but there are many more. Thinking, for example, in the classification problem, many times we would be interested in finding a trade-off between accuracy and recall (proportion of relevant instances retrieved) or precision (proportion of the retrieved instances that are relevant), depending on the problem we are working on. Another example is the trade-off between the accuracy and complexity of the models. Which is better, to use a combination of simple models (ensemble) or a very complex one (e.g., deep learning)? There is no clear answer.

Finally, we have seen that big data concerns different areas of knowledge, so we believe that, to provide high quality and complete solutions to these problems, the collaboration of experts from different areas in multidisciplinary teams will be necessary. In other words, paraphrasing Donoho, ``expand our boundaries beyond the classical domain of  statistics''\cite{donoho2015}. This would lead us to a new discipline that we may call {\it data learning}.







\bibliography{mybibfile}
\end{document}